\documentclass[lettersize,journal]{IEEEtran}
\usepackage{amsmath,amsfonts}
\usepackage{algorithmic}
\usepackage{algorithm}
\usepackage{array}
\usepackage[caption=false,font=normalsize,labelfont=sf,textfont=sf]{subfig}
\usepackage{textcomp}
\usepackage{stfloats}
\usepackage{url}
\usepackage{verbatim}
\usepackage{graphicx}
\usepackage{cite}
 \usepackage{makecell}

\usepackage{url,multirow,array,amsfonts,amssymb,array,cite,amsmath,amsthm,picinpar,color,soul,colortbl,graphicx,bm}
\usepackage[justification=centering]{caption}
\begin{document}

\title{Towards Intelligent Transportation with Pedestrians and Vehicles In-the-Loop: A Surveillance Video-Assisted Federated Digital Twin Framework}

\author{Xiaolong~Li, Jianhao Wei, Haidong Wang, Li Dong, Ruoyang Chen, Changyan Yi, ~\IEEEmembership{Senior Member,~IEEE} \\ Jun Cai,~\IEEEmembership{Senior Member,~IEEE}, Dusit Niyato,~\IEEEmembership{Fellow,~IEEE}, Xuemin (Sherman) Shen,~\IEEEmembership{Fellow,~IEEE}
\IEEEcompsocitemizethanks{\IEEEcompsocthanksitem X. Li, J. Wei, H. Wang, and L. Dong are with the Xiangjiang Laboratory, Hunan Provincial General University Key Laboratory of IoT Intelligent Sensing and Distributed Collaborative Optimization,  Hunan University of Technology and Business, Changsha, 410205, China (Email: lxl@hutb.edu.cn, jianhao@hnu.edu.cn, whd@hutb.edu.cn, Dlj2017@hunnu.edu.cn).  \protect
\IEEEcompsocthanksitem  R. Chen and C. Yi are with the College of Computer Science and Technology, Nanjing University of Aeronautics and Astronautics, China (Email: ruoyangchen@nuaa.edu.cn, changyan.yi@nuaa.edu.cn). \protect
\IEEEcompsocthanksitem J. Cai is with the Department of Electrical and Computer Engineering, University of Concordia, Montreal, Canada. (Email: Jun.cai@concordia.ca).  \protect
\IEEEcompsocthanksitem  D. Niyato is with the College of Computing and Data Science, Nanyang Technological University, Singapore (Email: dniyato@ntu.edu.sg). \protect
\IEEEcompsocthanksitem   X. Shen is with the Department of Electrical and Computer Engineering, University of Waterloo, Canada (Email: sshen@uwaterloo.ca).
}
}

% The paper headers
\markboth{}%
{Shell \MakeLowercase{\textit{et al.}}: A Sample Article Using IEEEtran.cls for IEEE Journals}

\maketitle

\begin{abstract}
In intelligent transportation systems (ITSs), incorporating pedestrians and vehicles in-the-loop is crucial for developing realistic and safe traffic management solutions. However,
there is falls short of simulating complex real-world ITS scenarios, primarily due to the lack of a digital twin implementation framework for characterizing interactions between
pedestrians and vehicles at different locations in different traffic environments. In this article, we propose a surveillance video assisted federated digital twin (SV-FDT) framework to empower ITSs with pedestrians and vehicles in-the-loop. Specifically, SV-FDT builds comprehensive pedestrian-vehicle interaction models by leveraging multi-source traffic surveillance videos. Its architecture consists of three layers: (i) the end layer, which collects traffic surveillance videos from multiple sources; (ii) the edge layer, responsible for semantic segmentation-based visual understanding, twin agent-based interaction modeling, and local digital twin system (LDTS) creation in local regions; and (iii) the
cloud layer, which integrates LDTSs across different regions to construct a global DT model in realtime. We analyze key design requirements and challenges and present core guidelines for SVFDT’s system implementation. A testbed evaluation demonstrates its effectiveness in optimizing traffic management. Comparisons with traditional terminal-server frameworks highlight SV-FDT’s advantages in mirroring delays, recognition accuracy, and subjective evaluation. Finally, we identify some open challenges and discuss future research directions.
\end{abstract}

\section{Introduction}
Intelligent transportation systems (ITSs) are envisioned to utilize information and communication technologies to enhance road safety and traffic management \cite{HChen}. Since pedestrians and vehicles are the natural and primary participants in transportation systems, incorporating both entities into real-time decision-making and control processes becomes necessary for ITSs, giving rise to the concept of pedestrian and vehicle in-the-loop. Such an “in-the-loop” scheme requires to continuously collect and process data from various sources, dynamically adjusting traffic operations based on real-time pedestrian-vehicle interactions, with both vehicles and pedestrians acting as active participants and decision-makers in the traffic management process. Compared to traditional ITSs, which primarily focus on vehicle modeling and often neglect pedestrians and their impacts  \cite{TYu, LYang, LKhan},  ITSs with pedestrians and vehicles in the loop can model and analyze the mutual influence between vehicles and pedestrians to provide safer, more efficient, and human-satisfying traffic management. However, enabling ITS applications with pedestrians and vehicles in-the-loop is non-trivial. It requires a cohesive system to characterize real-world traffic environments and conduct behavioral simulations of pedestrians and vehicles.

Recently, digital twin (DT) has been increasingly recognized as a powerful tool for building ITSs, offering real-time virtual representations of traffic operations to support data-driven decision-making and optimize traffic management \cite{LNie}. The federated digital twin (FDT), which can be seen as an advanced form of DT \cite{LKhan}\cite{TYu}, allows the integration of geographically dispersed DTs into a unified framework. FDT offers significant potential for ITSs to achieve pedestrians and vehicles in-the-loop, even when traffic participants, environments, and infrastructures are spread across different regions, each with numerous features and data sources. In addition, FDT allows data processing to be distributed, and thus can improve scalability and reduce latency. Furthermore, FDT enhances ITS' ability to predict future events using real-time data updates from multiple sources. All these capabilities are vital for proactive decision-making, such as pedestrian crossing safety and traffic signal optimization over a large area with multiple correlated intersections.

To achieve comprehensive and accurate FDT modeling, data from different sources are needed. Traffic surveillance cameras, as a primary component of ITS,  offer a cost-effective solution for large-scale transportation system monitoring. Currently, over one billion traffic surveillance cameras are installed worldwide. These cameras, strategically positioned along roadsides and intersections, facilitate seamless monitoring of entire intersections, crosswalks, and extended roadway segments. Unlike other terminal devices, such as radars or in-vehicle sensors, traffic surveillance cameras can provide continuous and real-time surveillance videos, which offer rich contextual information while accurately capturing the movements and interactions of pedestrians and vehicles. Thus, integrating real-time surveillance videos becomes essential for the effective functioning of FDTs in ITS, as they support fine-grained DT visualization, accurate pedestrian-vehicle interaction modeling, and real-time synchronization of FDT models with dynamic traffic conditions. Despite its significance, the potential of surveillance videos remains unexplored in ITSs.

In recent years, several DT frameworks have been developed to advance ITS applications \cite{LKhan, TYu, ZWang, ZOWang, LYang}. For example, Wang {\ et al.} in \cite{ZWang} proposed a cloud-edge-device collaborative DT framework for mobility services. However, the implementation details of the DT were not thoroughly addressed. A DT co-simulation framework for pedestrians and vehicles was introduced in \cite{ZOWang}, but it did not involve constructing twin agent models for these traffic participants. Khan {\it et al.} in \cite{LKhan} and Yu {\it et al.} in \cite{TYu} proposed two general FDT architectures for ITSs, yet neither study provided a technical implementation of the FDT system. Yang {\it et al.} in \cite{LYang} presented an FDT-enabled collision warning framework for autonomous driving, focusing on a semi-asynchronous federated learning scheme to reduce training delay. However, all these works primarily focus on high-level architectural designs or specific applications, lacking a comprehensive framework that integrates modeling the pedestrian-vehicle interactions and real-time FDT representations in complex traffic environments.

To address these limitations, in this article, we propose a surveillance video-assisted FDT implementation framework (SV-FDT) for ITSs with pedestrians and vehicles in-the-loop. The proposed SV-FDT is built on a cloud-edge-end collaborative architecture and introduces the agent concept to construct pedestrian-vehicle interaction models. The framework integrates semantic segmentation technology \cite{JTong}\cite{YZhao} for processing surveillance videos, which allows pix-level object recognition and offers granular insights into video data. SV-FDT can achieve timely video acquisition and data integration from multiple sources, precise modeling of pedestrian-vehicle twin agents at different positions, and seamless aggregation of local DT models across different regions. To our best knowledge, SV-FDT is the first DT implementation framework to enable ITSs with pedestrian and vehicle in-the-loop. Our key contributions are summarized as follows:

\begin{itemize}
\item We propose SV-FDT, a novel cloud-edge-end collaborative framework for implementing ITSs with pedestrians and vehicles in-the-loop. Its architecture includes i) an end layer that collects surveillance videos from widespread traffic surveillance cameras, ii) an edge layer that performs visual understanding, agent-based interaction modeling, and LDTS creation in local regions, and iii) a cloud layer that integrates LDTSs across geographically dispersed regions to construct a global DT model in real-time. To unlock its full potential, SV-FDT seamlessly integrates algorithms for semantic segmentation, semantic-to-code transformation, and agent-based modeling.
\item We analyze key design requirements and challenges for semantic segmentation-based FDT construction in SV-FDT, offering practical guidance for deploying FDTs via traffic video surveillance. We also highlight potential but promising solutions to these challenges, inspiring effective SV-FDT implementation in dynamic transportation.
\item We validate SV-FDT through a case study in the CARLA simulation environment to optimize traffic light control in traffic management. Our results show that SV-FDT outperforms traditional terminal-server frameworks in terms of mirroring delay, recognition accuracy, and subjective evaluations.
\item We summarize the open challenges in this field and propose future research directions.
\end{itemize}

\section{Framework of SV-FDT}
\label{sec:ITS_system}
\subsection{System Architecture and Key Techniques}
The proposed SV-FDT system architecture consists of three layers: end, edge, and cloud, as illustrated in Fig. 1.

\begin{figure*}[!t]
	\centering
	\includegraphics[width=0.75\textwidth]{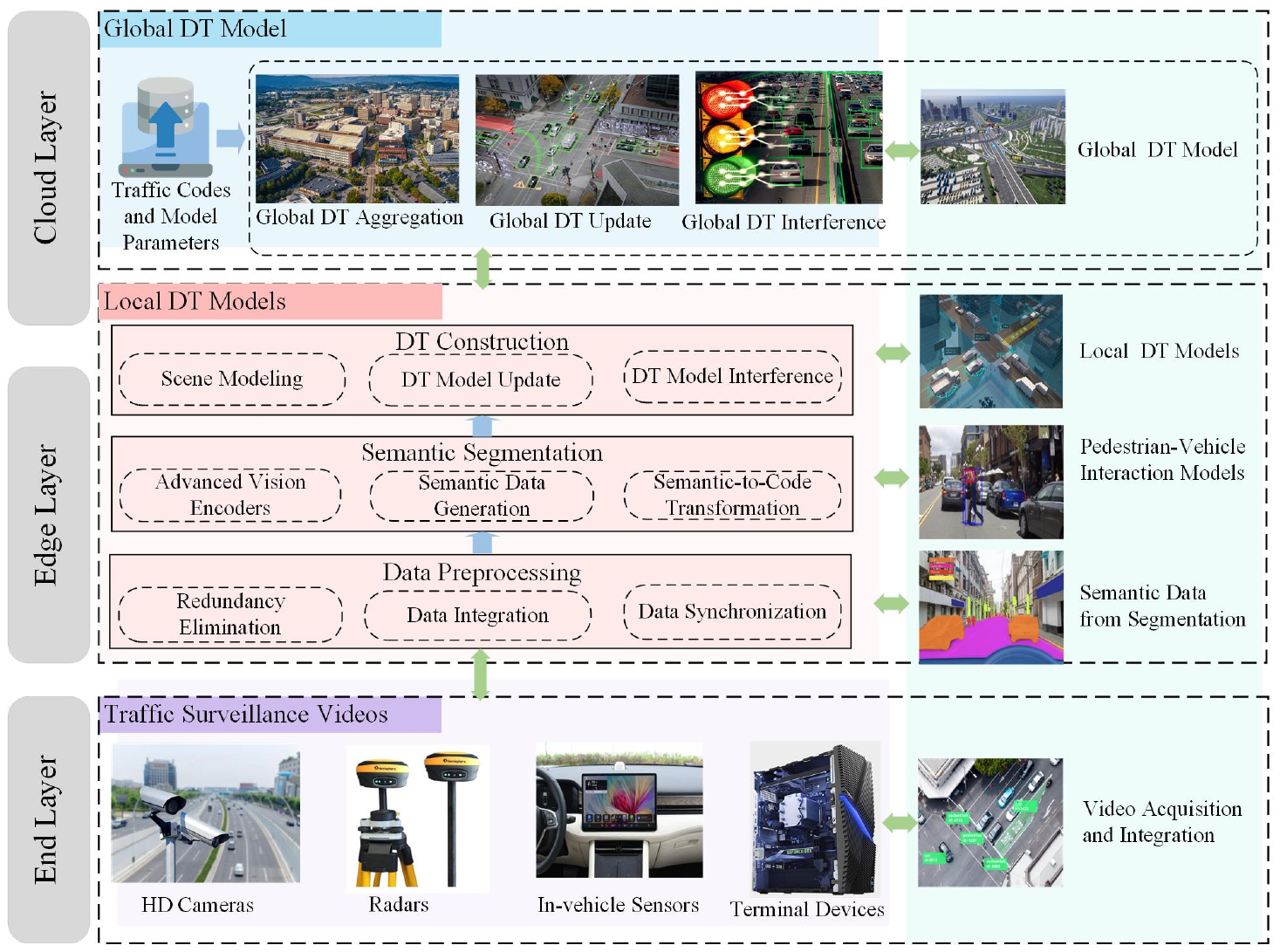}
	\vspace{-1.5em}
	\caption{The overall architecture of the proposed SV-FDT system. SV-FDT consists of three layers: the end layer contains multiple terminal devices; the edge layer handles tasks for data preprocessing, semantic segmentation, and local DT construction; and the cloud layer constructs the global model and conducts model inferences for ITS applications.}
\end{figure*}

\textbf{End Layer:} This layer consists of terminal devices such as surveillance cameras, radars, and in-vehicle sensors. Its primary function is to collect large-scale, real-time traffic data, particularly surveillance videos. For instance, high-definition (HD) surveillance cameras capture detailed visual data on traffic scenarios, such as vehicle movement, pedestrian crossings, and road conditions. Radars detect and track vehicles and pedestrians, providing information on their speed, location, and movement. In-vehicle sensors, e.g., GPS and ultrasonic sensors, contribute precise real-time localization and movement data. In addition to real-time data collection, these devices with integrated processing modules perform data fusion. Combining information (e.g., the locations and speeds of vehicles and pedestrians) from multiple sources can reduce uncertainties from solely relying on a single data source.

\textbf{Edge Layer:} At the edge layer, edge nodes, equipped with GPUs or NPUs, are responsible for data preprocessing tasks,  including i) redundancy elimination to video data uploaded by terminal devices for abundant storage and bandwidth; ii) data aggregation to obtain unified data from heterogeneous sources, e.g., traffic cameras, radars, and in-vehicle sensors, by combining radar measurements and in-vehicle sensor data with the visual analysis results; and iii) synchronization of temporal data to minimize latency discrepancies \cite{JChen}. More importantly, the edge nodes should conduct semantic segmentation to classify each pixel within each frame of surveillance video, enabling effective identification of pedestrians, vehicles, road markings, traffic signs, barriers, and weather conditions. Powered by advanced vision encoders, it extracts detailed visual features for precise pixel-level classification, continuously generating structured semantic data that accurately represents current traffic scenes. After that, the edge layer adopts a semantic-to-code transformation module that converts semantic data into machine-readable traffic codes, enabling their direct use in traffic simulation systems and facilitating the DT creation and update of physical environments by edge nodes. Meanwhile, the edge nodes transmit traffic codes and model parameters of twin agents to the cloud layer for aggregation to maintain spatiotemporal consistency between local and global DT models. Decision-making and inference results are returned to end-layer terminal devices, enabling adaptive signal control, path optimization, and personalized guidance to improve pedestrian safety and vehicle efficiency.

\textbf{Cloud Layer:} The cloud layer aggregates local DT models from distributed edge devices, creating an up-to-date global DT representation using traffic codes and agent-based model parameters across the transportation network. The global DT model continuously adapts to updates from local models, capturing new patterns in pedestrian-vehicle interactions. Only traffic codes and model parameters are transmitted to protect data privacy, safeguarding sensitive information about pedestrians and vehicles. After updating the global model, the cloud layer distributes global DT model parameters to the edge layer for synchronized updates of local models, enhancing decision-making accuracy.

\subsection{Example Scenarios}
Here, we present three typical application scenarios of the proposed SV-FDT framework for ITSs with pedestrians and vehicles in-the-loop.

\textbf{Extreme Pedestrian-Vehicle Flow Testbed In-the-Loop:}
SV-FDT can effectively implement ITSs with pedestrians and vehicles in-the-loop by providing a dynamic testbed to manage their interactions. The framework can support various automated driving levels, from Level 0 (fully manual) to Level 5 (fully autonomous). In each scenario, SV-FDT simulates real-world traffic challenges, such as sudden pattern changes and adverse weather conditions, and adapts by incorporating real-time data from diverse sources. This integration allows the testbed to reflect the varying capabilities of autonomous systems across different levels. By leveraging predictive analytics and collaborative decision-making, FDTs proactively identify and manage potential pedestrian-vehicle conflicts, unlike traditional testbeds that can only react after these situations occur.

\textbf{Emergency and Disaster Management:}
Integrating pedestrians and vehicles into SV-FDT can strengthen emergency and disaster management within ITSs. Traditional DT systems without in-the-loop characteristics rely on historical traffic data or static models for delayed emergency responses. Unlike these systems,  SV-FDT can continuously monitor and simulate real-world traffic scenarios, and adapt vehicles to unexpected traffic events and situations, including traffic incidents, road blockages, vehicle breakdowns, and sudden pedestrian crowds. Additionally, the system uses predictive models to anticipate emergencies, enabling the accurate assessment of traffic management strategies such as rerouting vehicles and safety notification of pedestrians. The in-the-loop characteristic means that the behaviors of pedestrians and vehicles are integrated into the system's decision-making processes, allowing for proactive rather than reactive management.

\textbf{Customized Navigation and Travel Guidance:}
Without in-the-loop FDTs, conventional navigation guidance systems rely on static or historical traffic data. However, they do not account for real-time conditions such as pedestrian activities, sudden traffic disruptions, or evolving road conditions. SV-FDT can be instrumental in enabling personalized navigation systems that cater to the specific needs of both drivers and pedestrians. By integrating data from various sources, such as surveillance cameras and in-vehicle sensors, FDTs create a dynamic, user-centered environment that adapts to individual preferences. For example, commercial drivers can receive guidance to avoid narrow streets or select routes suitable for heavy vehicles. Likewise, family drivers are provided with optimized routes that prioritize safety and efficiency. This adaptability allows FDTs to provide personalized, context-aware travel guidance that enhances the user experience and facilitates seamless interactions between pedestrians and vehicles in ITSs.

\subsection{Design Requirements and Challenges}
To establish a cloud-edge-end collaborative framework for real-time FDT construction in ITS, we identify the following key design requirements and challenges:

\subsubsection{Ultra-Reliable and Agile Global DT Model Aggregation with Pedestrian-Vehicle Interactions}
Real-time pedestrian-vehicle interactions require ultra-reliable data transmission to ensure safety and enable quick adjustments to traffic signals and vehicle routing. Therefore, achieving ``ultra-reliability'' and ``agility'' in aggregating global DT models from local data is paramount.

\textbf{Timely Video Acquisition and Integration with Precise Co-Camera Control:} Real-time video data collection and integration from potentially heterogeneous terminal cameras present challenges in synchronization and coordination \cite{YYang}. Therefore, a distributed network protocol is essential for timely video acquisition and integration, ensuring that video data is aligned temporally and processed without delays for real-time and accurate DT modeling \cite{XZhou}. Coordination techniques, such as distributed scheduling and synchronization, facilitate seamless communication and data sharing among devices, ensuring efficient real-time video data aggregation from multiple cameras. To ensure smooth HD streaming, the end devices, typically the surveillance camera, must achieve a minimum video transmission rate of 10 Mbps and a frame rate of 30 frames per second, supporting H.264 or H.265 video coding. Besides, cameras can be retrofitted with GPU-powered hardware modules to improve efficiency by using lightweight semantic segmentation algorithms to extract granular motion features of traffic elements (e.g., speeds, locations, and directions). Driven by these technologies, co-camera control enables multiple cameras to collaboratively capture videos from various angles and locations, ensuring comprehensive data integrity.

\textbf{Multi-Source Data Processing and Distributed Data Synchronization with Privacy Protection:} Leveraging multimodal data collected from diverse sources in ITS, e.g., HD cameras and radars, provides valuable insights into ITS' full feature spaces, including the appearance, locations, directions, and speed of both pedestrians and vehicles. Consolidating such information requires advanced multimodal data processing and distributed synchronization techniques to create unified feature profiles. This data processing and synchronization process must occur efficiently, ensuring quick updates to the global DT model. However, collecting and transmitting multimodal data across numerous sources may face significant privacy challenges. From a cross-layer security perspective, privacy-aware access control and signal protection in both the physical and data link layer, user indistinguishability enhancement, cross-device authentication, and customized packet encryption in upper layers are essential. Advanced technologies, such as physical layer security, local differential privacy, zero-trust architecture, and blockchain, can effectively mitigate these risks.

\textbf{Elastic Resource Orchestration for Fine-Grained DT Visualization under Traffic Dynamics:} Edge servers play a critical role in constructing DTs with low-latency requirements yet face significant challenges. High latency can hinder the real-time monitoring of pedestrians and vehicles, leading to uncoordinated interactions and safety risks. Therefore, an integrated approach to task scheduling, communication spectrum management, computing resources, and storage allocation becomes essential \cite{LDong}. Improving data transmission reliability requires employing ultra-reliable, low-latency communication (URLLC) technologies such as time-sensitive networks and mobile edge computing. These technologies support DT services, including immersive virtual reality and collaborative autonomous driving. For instance, immersive virtual reality demands high-rate video delivery from end to edge, requiring 99.999999$\%$  reliability and a latency of no more than 1 ms \cite{KSKim}.

\subsubsection{Customized and Sustainable  Multi-Agent Maintenance}
To enable customized pedestrian-vehicle interactions in ITSs, traffic participants are modeled as independent twin agents, containing virtual models replicating the physical objects and artificial intelligence for predictive decision-making. ``Customization'' addresses diverse and complex user needs, while ``sustainability'' focuses on the long-term evolution of FDT systems. Computational optimization is crucial for achieving customized and sustainable multi-agent maintenance, ensuring that FDT systems adapt to user needs and support sustainable development over time.

\textbf{Achieving Lightweight and Redundant-Free Replication of Large-Scale Video Footage:} In DTs, videos uploaded by multiple terminals often contain significant redundancy, consuming excessive storage space and network bandwidth while increasing the risk of exposing sensitive pedestrian-vehicle information. Redundancy can hinder semantic segmentation, reducing the responsiveness and safety of pedestrian-vehicle interactions. Therefore, de-redundancy algorithms are essential for constructing DTs and reducing storage consumption. To obtain lightweight and redundant-free replication, techniques such as data deduplication algorithms, model pruning, knowledge distillation, video compression, and semantic communication can be employed, promoting effective management of both pedestrians and vehicles.

\textbf{Balancing Standardized and Personalized Model Extraction for Pedestrian-Vehicle Agents:} Achieving a balance between standardization and personalization is crucial in pedestrian-vehicle interactions. Standardized agent models should support some industry standards such as UL-4600\footnote{UL-4600 is available at https://www.intertek.com/automotive/ul-4600.}  and can capture general features, such as typical pedestrian behavior patterns and vehicle operation modes. In contrast, personalized agent models focus on individual differences, including walking patterns, driving styles, and vehicle configurations. Although commonalities exist,  significant individual variations must be considered to model pedestrian-vehicle interactions precisely. Standardized models depend on extensive data collection, whereas personalized models require individual feature extraction via deep neural networks. An imbalance between the standardized and personalized DT model construction in ITS may result in low-quality global and local models within FDT construction. This significantly degrades the accuracy of both i) the global decision-making, e.g., traffic control and emergency management; and ii) the local decision-making, e.g., customized navigation and travel planning.

\textbf{Forward-Thinking and Spatially Versatile Cross-Model Migration under Ubiquitous Mobility:} Predicting user behaviors in DT models for highly mobile users, such as pedestrians and drivers, is challenging due to its inherent unpredictability. For instance, drivers might deviate from suggested routes due to personal preferences or unforeseen circumstances, complicating the transfer of models. To tackle this issue, generative AI (GAI) technology can simulate potential user behaviors and decision-making paths, aiding in scenario forecasting. Transfer learning techniques can also be employed to personalize and refine the model, enhancing its adaptability to individual user needs. This approach facilitates effective management of interactions between pedestrians and vehicles, ultimately delivering more intelligent and tailored services.

\section{SYSTEM IMPLEMENTATION}
The SV-FDT framework presented in Section II shows that cloud-edge-end collaboration is critical for real-time data integration and distributed processing in dynamic traffic environments, and continuous model refinement and data management can ensure local responsiveness and global coherence. In this section, we describe in detail the implementation of the proposed SV-FDT framework.

\begin{figure*}[!t]
	\centering
	\includegraphics[width=0.6\textwidth]{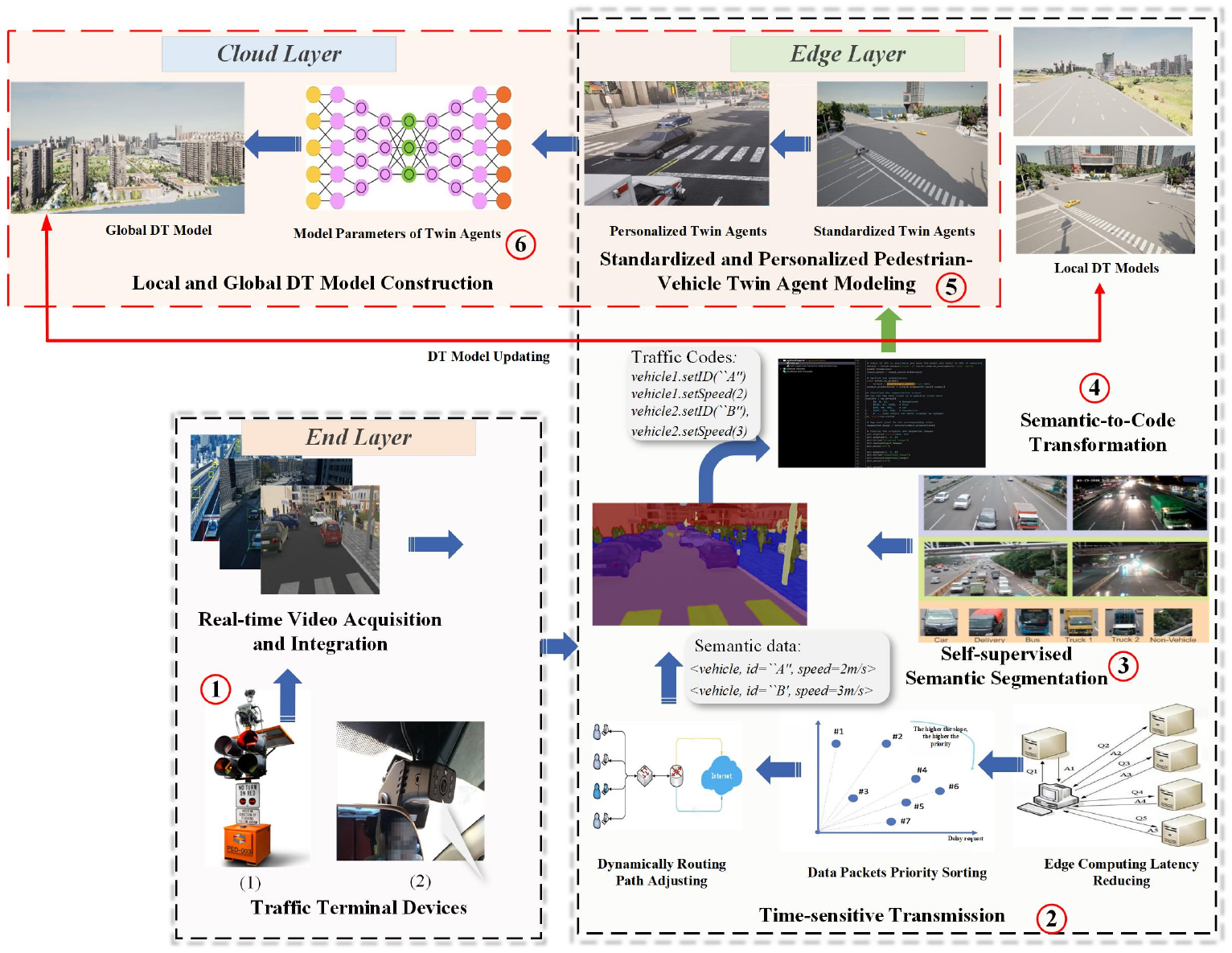}
	\caption{Systematic framework of the proposed SV-FDT. The semantic segmentation algorithm generates semantic data for vehicles: {\it $<$vehicle, id=``A'', speed=2m/s$>$, $<$vehicle, id=``B'', speed=3m/s$>$}. The semantic-to-code transformation module converts the semantic data into the following traffic codes: {\it vehicle1.setID(``A''), vehicle1.setSpeed(2), vehicle2.setID(``B''), vehicle2.setSpeed(3)}. Local and global DT models then transmit and execute these codes.}
\end{figure*}

\subsection{DT Construction by Cloud-Edge-End Collaboration}
Fig. 2 illustrates an implementation of the SV-FDT framework, which incorporates six key steps: real-time video acquisition and integration, time-sensitive transmission, self-supervised semantic segmentation, standardized and personalized pedestrian-vehicle twin agent modeling, and local and global DT model construction. SV-FDT leverages CARLA simulation environments to create traffic scenes and develop local and global DT models. CARLA, an open-source simulator built on Unreal Engine, generates high-fidelity traffic environments and accurately simulates traffic participant behaviors. SV-FDT leverages CARLA's map editor to design immersive traffic scenarios with vehicles, pedestrians, and detailed surroundings.

\textbf{At the end layer}, SV-FDT employs HD cameras and other terminal devices to collect and integrate real-time surveillance videos, establishing a foundation for constructing local and global DT models.

\textbf{At the edge layer}, before semantic segmentation, edge nodes perform essential data preprocessing, including redundancy elimination, data integration, and synchronization. After that, semantic segmentation algorithms are applied to identify traffic elements in the video, such as pedestrians, vehicles, road markings, traffic signs, barriers, and weather conditions. Both traffic elements' static attributes (e.g., appearance, location, and direction) and dynamic attributes (e.g., speed, behaviors, and movement trajectories) can be captured. Based on these attributes, semantic segmentation generates continuous, structured semantic data to support a granular understanding of visual inputs. The semantic-to-code transformation module then translates semantic data into executable traffic codes, which are used to generate digital twinning of real-world scenes. Such traffic codes define attributes of vehicles and pedestrians, including locations, speeds, and movement directions. Edge nodes continuously update local DT models and simulate pedestrian and vehicle behaviors by successively executing generated traffic codes.

\textbf{At the cloud layer}, cloud servers constantly receive semantic data and model parameters of pedestrian-vehicle agents from edge nodes to achieve a unified global DT representation while ensuring spatiotemporal coherence without transmitting sensitive visual data \cite{ZWu}. We could exploit the global DT model to develop and implement ITS applications running on cloud servers, such as traffic signal control and traffic flow prediction.

\subsection{Implementation of Local and Global Digital Twins}
\textbf{Real-time Video Acquisition and Integration:} Video compression standards, such as H.264 and H.265, are applied to reduce bandwidth consumption while maintaining high video quality. Adaptive bitrate streaming adjusts video quality dynamically according to bandwidth availability, ensuring seamless streaming. With the time synchronization protocol, all video frames and sensor data are timestamped for accurate synchronization during analysis. Multi-sensor fusion algorithms, such as Kalman Filters, are used to integrate data from various sources to ensure synchronized object representation across modalities.

\textbf{Time-sensitive Transmission:} URLLC technologies are used to enable immersive interactions in DTs. Edge computing minimizes latency by positioning resources closer to end users, enabling parallel processing of multimodal data for seamless pedestrian-vehicle interactions. We prioritize data packets such as collision warnings based on latency requirements and importance, ensuring that critical data packets can be transmitted first. Additionally, we dynamically adjust the transmission rate and route paths of data streams according to real-time network conditions and transmission needs.

\textbf{Self-supervised Semantic Segmentation:} We prefer self-supervised semantic segmentation over unsupervised methods for its ability to deliver real-time, pixel-level insights into traffic elements without manual labeling, which is time-consuming and costly. Advanced models such as  DINO\footnote{https://github.com/IDEA-Research/MaskDINO} excel in capturing long-range dependencies and contextual relationships, making them ideal for complex traffic scenarios. Granularity-controllable segmentation allows precise customization of output detail to align with user requirements \cite{YZhao}. Techniques such as contrastive learning enhance the model's ability to identify key attributes at the pixel level. For example, optical flow measures pedestrian and vehicle speeds, while pixel-level annotations precisely position traffic elements. By extracting static and dynamic attributes through semantic segmentation, we generate continuous, structured semantic data to support a detailed visual understanding of traffic scenes. This semantic data is converted into traffic codes via the semantic-to-code transformation module.

\textbf{Semantic-to-Code Transformation:} The semantic-to-code transformation module is a core component of the framework, converting structured semantic data into executable traffic codes for CARLA simulation. This process uses rule-based algorithms and advanced deep learning techniques, e.g., generative AI and large language models (LLMs), to generate traffic codes based on traffic elements and their attributes, as illustrated in Fig. 2. The transformation logic ensures accurate scene replication within CARLA environments by adhering to CARLA's APIs and command structures. We continuously monitor traffic code performance in CARLA to maintain accuracy, refining the transformation logic as needed to correct discrepancies. This iterative process ensures that the traffic codes are both precise and reliable.

\textbf{Standardized and Personalized Pedestrian-Vehicle Twin Agent Modeling:}  We develop pedestrian-vehicle twin agents with both standardized and personalized features to simulate realistic interactions. Using model-driven agent modeling, we treat traffic participants as independent, autonomous agents, creating dynamics that combine standardized traits for predictable behaviors in typical scenarios with personalized features based on individual-specific data. Each twin agent operates autonomously in the CARLA environment and independently perceives its surroundings, makes decisions, and executes actions. For instance, vehicles can adjust speed and trajectory in response to weather conditions or interactions with pedestrians and other vehicles. Agent movement is influenced by its attributes (e.g., direction and speed), interactions with other agents, the environment, and infrastructure. Standardized traits ensure common behaviors that align with typical traffic rules, while personalization allows agents to adapt to specific driver habits or pedestrian walking patterns. This hybrid approach strikes a balance between standardized and personalized behaviors, enabling agents to dynamically adjust and produce realistic, responsive interactions in complex traffic environments.

\textbf{Local and Global DT Model Construction:} Local and global DT models are built and updated using semantic data and model parameters of twin agents. Edge nodes construct 3D local panoramic DT models of roads, buildings, and infrastructure, either online or offline, and continuously update local DT models by executing traffic codes. These traffic codes dynamically adjust traffic element attributes, such as weather, vehicle positions, and pedestrian presence, keeping local DT models accurate and current. The global DT model aggregates semantic data and agent parameters from edge nodes, providing a unified, large-scale view across the transportation network. Data exchanges between edge nodes and cloud servers maintain privacy and spatiotemporal coherence by omitting sensitive visual information. As local  DT models update continuously, the global DT model synchronizes these updates to ensure a consistent, coherent representation of the transportation system across regions. Local DT models mirror specific traffic regions in ITS applications, e.g., navigation guidance, incorporating pedestrian-vehicle interactions and evolutionary reasoning with real-time updates. The global DT aggregates these updates, providing a comprehensive network view for global reasoning to guide vehicles and pedestrians along the most efficient or safest routes.

\section{Case Study and Experimental Results}
Given the overall architecture of the proposed SV-FDT framework, this section presents a pioneering case study on optimizing traffic light time setting, addressing design requirements and challenges as presented in Section II.

\begin{figure*}[!t]
	\centering
	\includegraphics[width=0.8\textwidth]{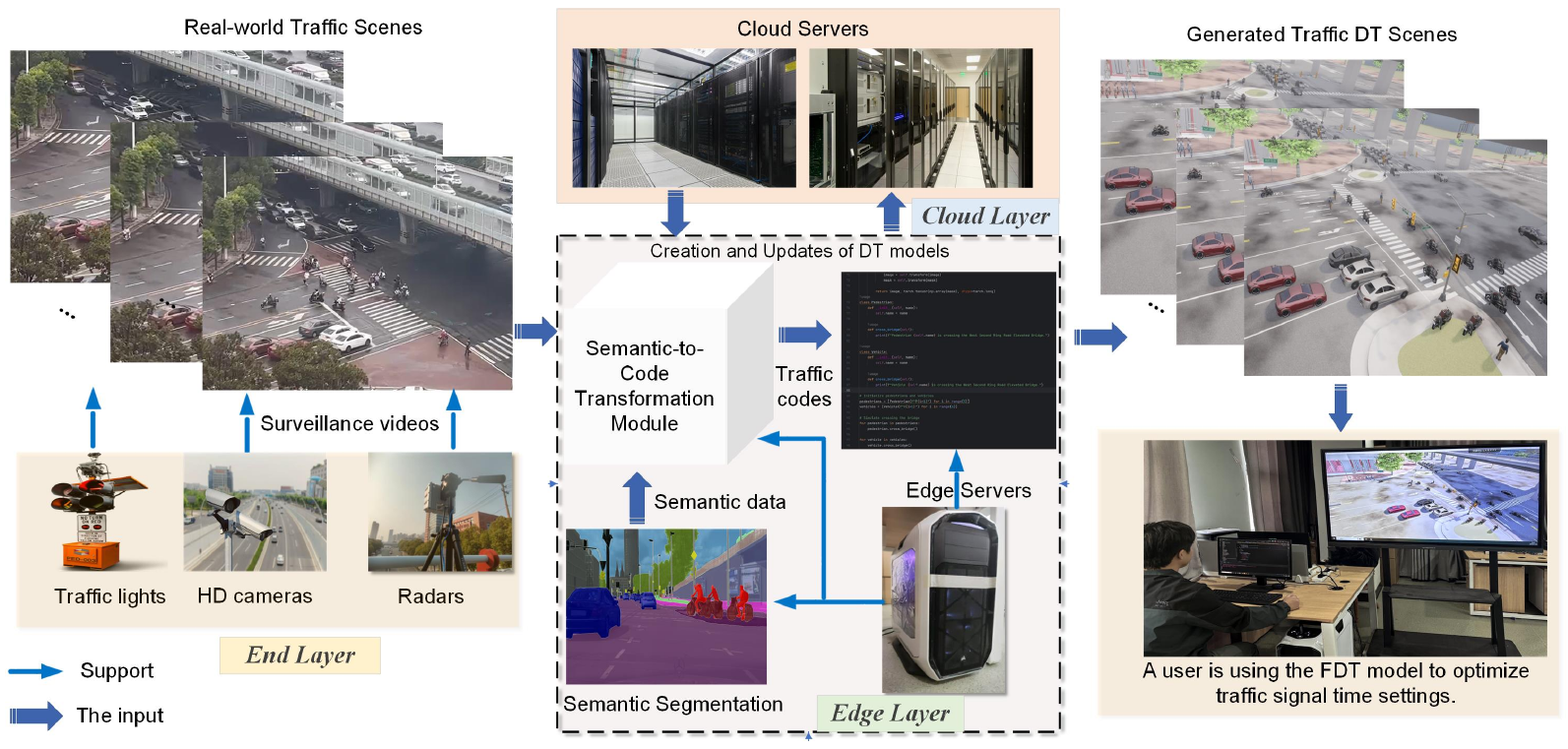}
	\vspace{-5em}
	\caption{The SV-FDT framework-based FDT testbed platform for optimizing traffic signal timing.}
	\label{fig:Fig3}
\end{figure*}
\vspace{-2em}

\subsection{Experimental Setup and System Design}
As illustrated in Fig. 3, we developed a testbed platform following SV-FDT to optimize traffic signal timing, a typical signal control environment in ITS. Terminal devices at the end layer include HD cameras, radars, and traffic lights. The HD cameras and radars collect visual data and can preprocess and integrate it. The edge layer comprises a desktop computer as the traffic signal control center and five laptops equipped with NVIDIA RTX 4060 GPUs and i9-14900HX processors, functioning as edge nodes. One edge node uses a lightweight semantic segmentation algorithm to monitor pedestrian and vehicle positions, generating continuous semantic data on their speeds and trajectories. Another edge node operates the semantic-to-code transformation module, converting semantic data into executable traffic codes for the CARLA simulation environment. The third edge node, dedicated to scene modeling, generates a 3D panoramic model of roads and surrounding buildings. The fourth and fifth edge nodes, two local DT modeling nodes, perform real-time updates of local DT models in two regions. The traffic signal control center uses the data of local DT models to optimize signal settings. Two cloud servers are deployed at the cloud layer: one generates the global DT model, while the other implements the global signal control optimization application. As shown in Fig. 4, SV-FDT constructs an FDT model for an area near the south campus of Hunan University of Technology and Business. The area features two intersections with different configurations, including crossroads and pedestrian crossings. Four pedestrian walkways intersect with vehicle lanes, and the traffic flow of pedestrians and vehicles varies significantly throughout the day. This makes the area an ideal and realistic urban traffic test environment. SV-FDT dynamically adjusts signal timings based on road width, pedestrian walking speed, and volume, enhancing road safety.

The experiments utilized 4-megapixel cameras capturing traffic with a bandwidth of 10 Mbps, supporting real-time data collection and processing at 30 frames per second. Drone aerial footage provided offline video data to the scene modeling node, enabling the pre-construction of a static 3D panoramic DT model. End nodes connect to edge nodes via Gigabit Ethernet for efficient video transfer, while edge-to-cloud communication occurs over 5G networks at speeds up to 200 Mbps. Lightweight UniMatch V2+ DINOv2 encoders \footnote{https://github.com/LiheYoung/UniMatch-V2}  perform semantic segmentation to track real-time pedestrian and vehicle flow, generating structured semantic data at 30 times per second sent to the semantic-to-code transformation node. This transformation node utilizes the WizardCoder language model as its foundational pretraining model. After fine-tuning with semantic data and traffic codes, WizardCoder generates executable traffic codes from structured semantic inputs. The local DT node loads the 3D panoramic model from the scene modeling node and uses CARLA to execute these codes, simulating real-world pedestrian-vehicle interactions at the frequency of 30.

Meanwhile, the global DT modeling node integrates semantic data from all local DT models, creating a comprehensive global DT model within CARLA. The aggregation and updates of the global DT are done less frequently than those of the local DTs. Instead, updates occur periodically every minute or when significant changes in traffic patterns, such as traffic events, are detected. Ultimately, SV-FDT dynamically adjusts signal timings to improve road safety. The case study showcases the feasibility and validity of the SV-FDT framework.

\begin{figure}[t]
\centering
\includegraphics[width=0.48\textwidth]{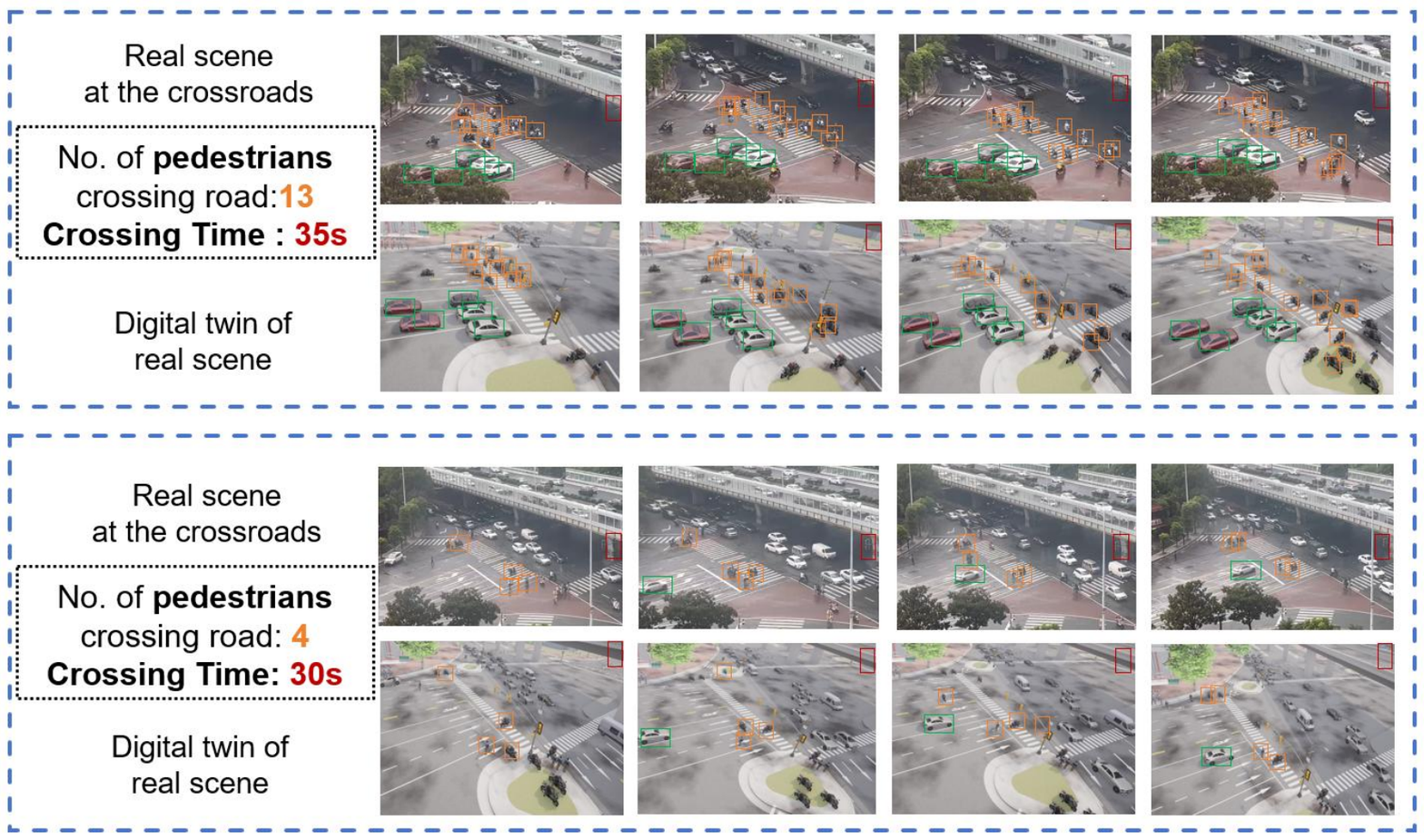}
\vspace{-2em}
\caption{Optimized traffic signal time settings based on road width, pedestrian walking speed, and volume.}
\label{fig:fig5}
\end{figure}

\subsection{Performance Evaluation}
Performance is evaluated on five metrics to validate the proposed platform: 1) \textbf{Mirroring delay}: The time required to create a digital twin model that mirrors its real-world counterpart, measured in milliseconds. 2) \textbf{Recognition accuracy}: The ratio of correct detections of pedestrians and vehicles to all positive detections, expressed as a percentage. 3$\&$4) \textbf{Traffic/Pedestrian flow}: The average number of vehicles/pedestrians passing a specific intersection within one minute. 5) \textbf{Subjective evaluations}: These are assessed through user feedback on video fluency, consistency between physical and cyber domains, and synchronized movements. As shown in Fig. 5(a), the SV-FDT framework outperforms the traditional terminal-server framework in DT modeling, achieving significantly lower average delay (AVE-DL), maximum delay (MAX-DL), minimum delay (MIN-DL), and jitter (JIT-DL). This is because, in the proposed SV-FDT, i) data processing occurs at the edge nodes closer to the data sources; ii) only redundancy-eliminated traffic codes and model parameters of twin agents rather than raw surveillance videos are sent to the cloud. Fig. 5(b) illustrates recognition accuracy, where SV-FDT also excels, demonstrating higher average (AVE-RA), maximum (MAX-RA), and minimum recognition accuracy (MIN-RA), along with lower jitter (JIT-RA). This is because semantic segmentation algorithms in SV-FDT are executed at the edge nodes that can continuously learn from local traffic, which guarantees overall recognition accuracy. Moreover, edge layer processing reduces latency, ensuring faster and more accurate recognition by enabling real-time data analysis without delays. Fig. 5(c)  and 5(d) illustrate traffic flow and pedestrian flow under four different pedestrian-vehicle density configurations. Both figures show traffic and pedestrian flow under four different pedestrian-vehicle density configurations. Compared to fixed traffic signal settings, the SV-FDT achieves notable performance gains using adaptive traffic signal timing. Finally, a subjective evaluation was conducted with ten volunteers testing the platform. Table I shows that the terminal-server framework struggles with DT modeling delays, discrepancies, and asynchronous movements, while the SV-FDT framework operates seamlessly. These findings confirm that our framework provides a superior and immersive Quality of Experience (QoE) in both objective and subjective measurements.

\begin{figure}[!t]
		\centering
			\includegraphics*[width=1.0\linewidth]{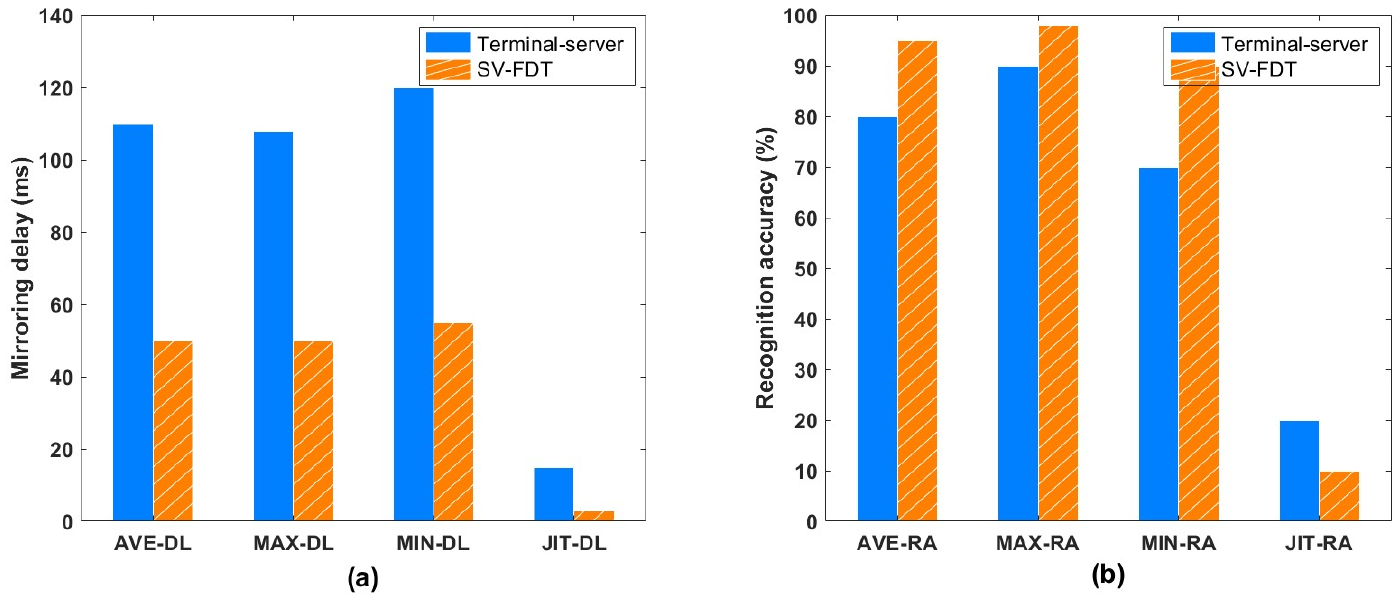} \label{subfig1}\\
			\vspace{-9em}
			\includegraphics*[width=1.0\linewidth]{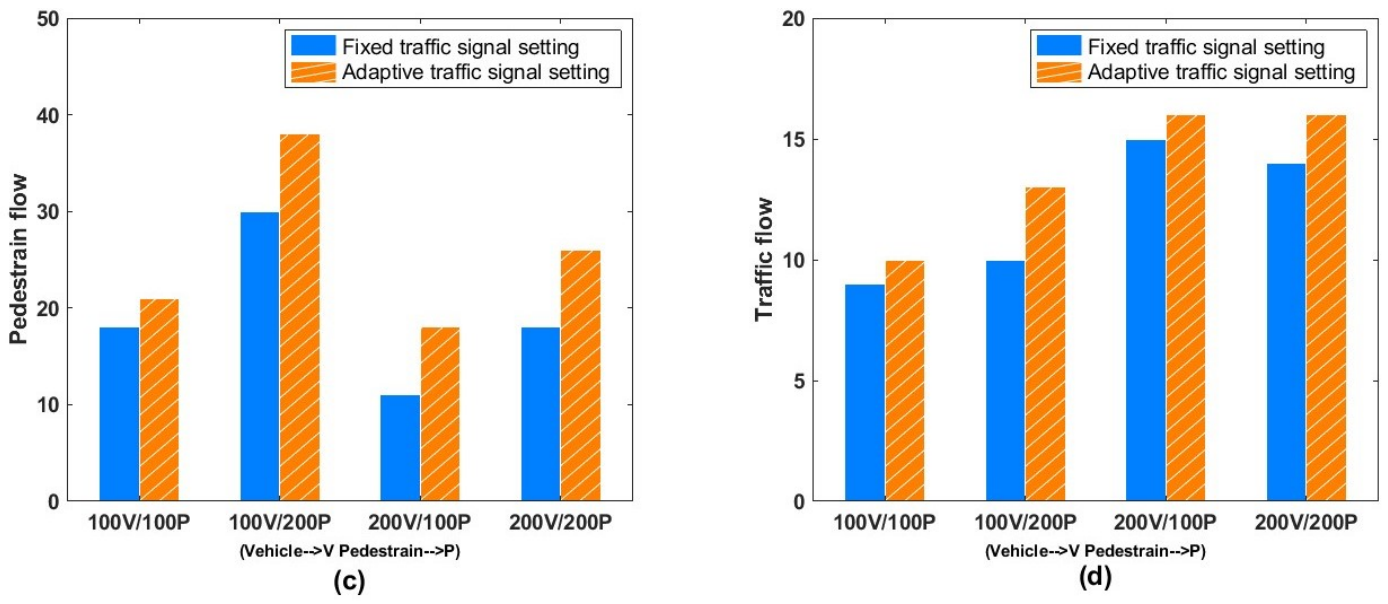} \label{subfig2}
			\vspace{-6em}
     \caption{Comparisons on objective measurements.}
	\end{figure}

\begin{table}[h]
\centering
\caption{Comparisons on subjective measurements.}
\begin{tabular}{c|c c c}
\Xhline{0.025cm}
\textbf{Framework} & \textbf{Video fluency} & \textbf{DT consistence} & \textbf{Sync.} \\ \hline
\Xhline{0.025cm}\hline
Terminal-server    & Choppy                & Discrepant                & Asynchronous        \\  \hline
SV-FDT     & Fluent                & Consistent                & Synchronous      \\  \hline
\Xhline{0.025cm}
\end{tabular}
\end{table}

\section{Conclusion and Future Research Directions}
This article presents a novel cloud-edge-end collaborative framework, SV-FDT, designed to revolutionize ITSs by integrating pedestrians and vehicles in the loop. The proposed framework includes a comprehensive system architecture and highlights key design requirements and challenges for semantic segmentation-based FDT construction. We demonstrate several potential applications, such as an extreme pedestrian-vehicle flow testbed and emergency and disaster management. A case study conducted in CARLA simulation environments validates the effectiveness of SV-FDT in optimizing traffic management. Our results show that SV-FDT outperforms traditional terminal-server frameworks in terms of mirroring delay, recognition accuracy, and subjective evaluations. Furthermore, we identify and outline several open challenges.

\begin{itemize}
\item \textit{High-Precision DT Extraction via Multimodal Fusion:}
Extracting and fusing features from multimodal traffic data (e.g., images, audio, text, videos) enhances the accuracy of FDT models in heterogeneous ITS environments. While traditional methods rely on single-modal data, multimodal models require real-time data acquisition and distributed storage, posing privacy and security challenges. Future work will focus on building FDT models using knowledge graphs and deep learning to improve security and accuracy.

\item \textit{Seamless DT Synchronization Across Diverse Vehicular Communication Systems:}
Vehicular communication devices vary in transmission rates, coverage, reliability, and interference. Real-time DT synchronization across these vehicular communication systems is challenging. Future efforts will leverage 6G technology for efficient wireless transmission and explore visible light communication (VLC), radio frequency identification (RFID), millimeter wave (mmWave), acoustic communication, and low Earth orbit (LEO) satellites to optimize DT synchronization.

\item \textit{Extending FDT to Future ITS with Hyper-Spatial Mobility:} As AI advances, future ITS will evolve into hyper-spatial systems for efficient, rapid, and secure transportation across rail, subway, drone, and autonomous vehicle networks. Future research will explore deep learning-based technologies and intelligent control systems to develop hyper-spatial FDT models, supporting autonomous decision-making, real-time route planning, state monitoring, performance evaluation, and fault prediction across modes.

\item \textit{Optimal Traffic Signal Control in Self-Evolving FDT Systems:}
Errors in DT models can severely impact traffic signal accuracy and decision-making. Achieving optimal FDT coordination is complex. Future work will focus on developing models to detect and correct inaccuracies in traffic DT systems, applying filtering techniques, and using federated learning and collaborative optimization to enhance data sharing and coordinated decision-making. Distributed and coordinated traffic signal control approaches based on multi-agent learning will be developed to respond to real-time traffic adaptively, improving flow and road safety.
\end{itemize}

\section*{Biography}

\small \textbf{Xiaolong Li (Member, IEEE)} is currently a Professor at Hunan University of Technology and Business in Changsha. His research interests include deep learning, intelligent transportation systems, and the Internet of Things.

\small \textbf{Jianhao Wei} is currently an associate professor at Hunan University of Technology and Business in Changsha. His research interests include deep learning, intelligent transportation systems, and the Internet of Things.

\small \textbf{Haidong Wang} is currently a lecturer at Hunan University of Technology and Business in Changsha. His research interests include deep learning, intelligent transportation systems, and digital twins.

\small \textbf{Li Dong} is currently an associate professor at Hunan University of Technology and Business in Changsha. Her research interests include deep learning, intelligent transportation systems, and digital twins.

\small \textbf{Ruoyang Chen} is currently pursuing a Ph.D degree at the College of Computer Science and Technology, Nanjing University of Aeronautics and Astronautics, China. His research interests include edge computing and digital twin.

\small \textbf{Changyan Yi (Senior Member, IEEE)} is a Professor at the College of Computer Science and Technology, Nanjing University of Aeronautics and Astronautics, China. His research interests include edge computing, industrial IoT, digital twin, 5G, and beyond.

\small \textbf{Jun Cai (Senior Member, IEEE)} is a Professor and PERFORM Centre Research Chair with the Department of Electrical and Computer Engineering, Concordia University, Canada. His research interests include edge/fog computing and eHealth.

\small \textbf{Dusit Niyato (Fellow, IEEE)} is a President's Chair Professor at the School of Computer Science and Engineering, Nanyang Technological University, Singapore. His research interests include edge intelligence, machine learning, and incentive mechanism design.

\small \textbf{Xuemin (Sherman) Shen (Fellow, IEEE)} is a University Professor at the Department of Electrical and Computer Engineering, University of Waterloo, Canada. His research focuses on network resource management, wireless network security, social networks, and vehicular ad hoc networks. He is a Canadian Academy of Engineering Fellow, a Royal Society of Canada Fellow, and a Chinese Academy of Engineering Foreign Fellow.


\begin{thebibliography}{1}
\bibliographystyle{IEEEtran}
\bibitem{HChen}
H. Chen, H. Kim, M. Ammous, G. Seco-Granados, G. Alexandropoulos, S. Valaee, and H. Wymeersch, ``RISs and sidelink communications in smart cities: the key to seamless localization and sensing,'' \textit{IEEE Commun. Mag.}, vol. 61, no. 8, pp. 140-146, Aug. 2023.

\bibitem{LKhan}
L. Khan, E. Mustafa, J. Shuja. F. Rehman, K. Bilal, Z. Han, and C. S. Hong, ``Federated learning for digital twin-based vehicular networks: architecture and challenges,'' \textit{IEEE Wireless Commun.}, vol. 31, no. 2, pp: 156-162, Apr. 2024.

\bibitem{TYu}
T. Yu, Z. Li, O. Hashash, K. Sakaguchi, W. Saad, and M. Debbah, ``Internet of federated digital twins: connecting twins beyond borders for society 5.0,'' \textit{IEEE IoTM.}, vol. 7, no. 5, pp. 74-81, Sep. 2024.

\bibitem{LYang}
L. Tang, M. Wen, Z. Shan, L. Li, Q. Liu, and Q. Chen, ``Digital twin-enabled efficient federated learning for collision warning in intelligent driving,'' \textit{IEEE Trans. Intell. Transp. Syst.}, vol. 25, no. 3, pp. 2573-2585, Mar. 2024.

\bibitem{LNie}
L. Nie, X. Wang, Q. Zhao, Z. Shang, L. Feng, and G. Li, ``Digital twin for transportation big data: a reinforcement learning-based network traffic prediction approach,'' \textit{IEEE Trans. Intell. Transp. Syst.}, vol. 25, no. 1, pp. 896-906, Jan. 2024.


\bibitem{ZWang}
Z. Wang, R. Gupta, K. Han, H. Wang, A. Ganlath, N. Ammar, and P. Tiwari, ``Mobility digital twin: concept, architecture, case study, and future challenges,'' \textit{IEEE Internet Things J.}, vol. 9, no. 18, pp. 17452-17467, Sep. 2022.

\bibitem{ZOWang}
Z. Wang, O. Zheng, L. Li, M. Abdel-Aty, C. Cruz-Neira, and Z. Islam, ``Towards next generation of pedestrian and connected vehicle in-the-loop research: a digital twin co-simulation framework,'' \textit{IEEE Trans. Intell. Veh.}, vol. 8, no. 4, pp. 2674-2683, Apr. 2023.


\bibitem{JTong}
J. Tong, Y. Zou, Y. Li, and R. Li, ``Lightweight frequency masker for cross-domain few-shot semantic segmentation,''  in \textit{Proc. of NeurIPS 2024}, Vancouver, BC, Canada, Dec. 2024.

\bibitem{YZhao}
Y. Zhao, K. Li, Z. Cheng, P. Qiao, X. Zheng,  R. Ji, C. Liu, L. Yuan, and J. Chen, ``GraCo: granularity-controllable interactive segmentation,'' in \textit{Proc. of CVPR 2024}, Seattle, WA, Jun. 2024, pp. 3501-3510.

\bibitem{JChen}
J. Chen, C. Yi, H. Du, D. Niyato, J. Kang, J. Cai, and X. Shen. ``A revolution of personalized healthcare: enabling human digital twin with mobile AIGC,'' \textit{IEEE Network}, Feb. 2024. doi:10.1109/MNET.2024.3366560.

\bibitem{YYang}
Y. Yang, Y. Shi, C. Yi, J. Cai, j. Kang,  D. Niyato, and X. Shen, ``Dynamic human digital twin deployment at the edge for task execution: a two-timescale accuracy-aware online optimization,'' \textit{IEEE Trans. Mob. Comput.}, vol. 23, no. 12, pp. 12262-12279, Dec. 2024.

\bibitem{XZhou}
J. Chen, C. Yi, S. D. Okegbile, J. Cai, and X. Shen, ``Networking architecture and key supporting technologies for human digital twin in personalized healthcare: a comprehensive survey,'' \textit{IEEE Commun. Surv. Tutor.}, vol. 26, no. 1, pp. 706-746, 1st. Quart. 2024.

\bibitem{LDong}
L. Dong, F. Jiang, M. Wang, Y. Peng, and X. Li, ``Deep progressive reinforcement learning-based flexible resource scheduling framework for IRS and UAV-assisted MEC system,'' \textit{IEEE Trans. Neural Networks Learn. Syst.}, Jan. 2024. doi: 10.1109/TNNLS.2023.3341067.

\bibitem{KSKim}
K. S. Kim, D. K. Kim, C.-B. Chae, S. Choi, Y. -C. Ko, J. Kim, Y.-G. Lim, M. Yang, S. Kim, B. Lim, K. Lee, and K. L. Ry, ``Ultrareliable and low-latency communication techniques for tactile internet services,'' \textit{Proc. IEEE}, vol. 107, no. 2, pp. 376-393, Feb. 2019.

\bibitem{ZWu}
Z. Wu, S. Sun, Y. Wang, M. Liu, B. Gao, Q. Pan, T. He, and X. Jiang, ``Agglomerative federated learning: empowering larger model training via end-edge-cloud collaboration,'' in \textit{Proc. of INFOCOM 2024}, Vancouver, BC, Canada, May. 2024, pp. 131-140.
\end{thebibliography}
\end{document}